\documentclass[twocolumn,showpacs,prl,aps,floatfix,reprint]{revtex4-1}

\usepackage{graphicx}
\usepackage{dcolumn}
\usepackage{bm}

\usepackage{hyperref}
\usepackage{amssymb}
\usepackage{braket}
\usepackage{graphicx}
\usepackage{dcolumn}
\usepackage{amsmath}
\usepackage[capitalize]{cleveref}
\usepackage{amsfonts}
\usepackage{color}
\usepackage{float}
\usepackage{booktabs}
\usepackage{diagbox}
\usepackage{multirow}
\usepackage{siunitx} 

\usepackage{placeins}
\usepackage{bm}
\begin{document}

\preprint{APS/123-QED}

\title{Dispersive optical model description of the CaFe experiment
}

\author{R. A. Ramon$^{1}$, M. C. Atkinson$^{2}$, and W. H. Dickhoff$^1$}

\affiliation{${}^1$Department of Physics,
Washington University, St. Louis, Missouri 63130}
\affiliation{${}^2$Department of Physics and Astronomy,
Haverford College, Haverford, PA 19041}

\date{\today}

\begin{abstract}
High-momentum ratios measured in the CaFe experiment using high-energy electron scattering off the CaFe nuclei $(i.e.~ {}^{40}\mathrm{Ca}, {}^{48}\mathrm{Ca}, {}^{54}\mathrm{Fe})$ have revealed a surprising correlation between the nuclear shell structure and the high-momentum components generated by short-range and tensor correlations (SRC). The non-local dispersive optical model (DOM) successfully describes the experimentally measured high-momentum ratios while simultaneously describing the nuclear bound-state properties and scattering observables for these CaFe nuclei. The experimentally measured average high-momentum ratio of ${}^{48}\mathrm{Ca}$/${}^{40}\mathrm{Ca}$ and ${}^{54}\mathrm{Fe}$/${}^{48}\mathrm{Ca}$ are $1.10~\pm~0.02$ and $1.49~\pm~0.03$, respectively. The corresponding DOM results are $1.05~\pm~0.05$ and $1.52~\pm~0.08$, which is the first successful description of these experimental results. 
The present DOM analysis of CaFe nuclei relied on an extended version of the DOM incorporating a shrinkable geometry to provide a more localized and stronger volume absorption at higher missing energies to correctly represent the SRC physics. The success of the DOM can be attributed to its simultaneous treatment of structure and reaction data, the latter requiring a depletion of the mean-field accompanied by the proper admixture of high-momentum components in the respective ground states.

\end{abstract}

\maketitle

A nucleus is a quantum many-body system, predominantly governed by the complicated nature of the strong nuclear force that holds the nucleons (protons and neutrons) together. The nuclear force is short-ranged containing strong attraction to overcome Coulomb repulsion but also sufficiently repulsive at shorter distances to prevent atomic nuclei from collapsing. Consequently, nucleons approaching very close to each other can momentarily form short-range and tensor correlated (SRC) pairs, primarily responsible for the high-momentum components observed in atomic nuclei~\cite{Hen:2017,Nguyen2026}. 
These SRC pairs are substantially localized with a separation of $ r\leq 1.0$ fm~\cite{miller:2020discovery} compared to the average nucleon separation of $r\approx1.7$ fm in nuclei~\cite{cruz:2021many}. These instantaneous correlated states cannot be described by the independent particle model (IPM), where nucleons move independently in a mean field with low momentum~\cite{Duer:2018}.

According to the IPM, the single-particle levels below the Fermi energy are fully occupied. This assumption is contradicted by $(e,e'p)$ experiments with nuclei ranging from ${}^{7}\mathrm{Li}$ to ${}^{208}\mathrm{Pb}$ that demonstrate that the occupancy of valence single-particle levels is depleted by $30-40\%$~\cite{Lapikas93, kelly:1996nucleon}. 
Modern calculations including long-range correlations (LRC) account for about a $20\%$~\cite{Dickhoff04} loss of strength near the Fermi surface, and SRC approximately cause a global depletion of $10-15\%$~\cite{Dickhoff04} for both valence and deeply bound states. These depleted strengths from the mean-field description are then partly compensated by high-momentum components extending to very high missing energy, resulting in a ``high-momentum tail'' in the momentum distribution of nucleons. Most of this high-momentum contribution comes from np SRC pairs, which represent the effect of the tensor force~\cite{Duer:2018}.

Recently, the CLAS Collaboration at Jefferson Lab used a 5.014 GeV electron beam on targets of ${}^{12}$C, ${}^{27}$Al, ${}^{56}$Fe, and ${}^{208}$Pb to investigate short-range dynamics and how neutron excess, $N/Z$, affects the high-momentum distribution. An increase in proton and a slight decrease in neutron high-momentum distributions were observed with increasing neutron excess~\cite{Duer:2018}.

In a more recent analysis of the CLAS collaboration, it is observed that the SRC pairing depends more on the nuclear shell structure than on the nuclear mass and the neutron excess~\cite{Nguyen2026}. A 10.549 GeV high-energy electron beam incident on ${}^{40}$Ca, ${}^{48}$Ca, and ${}^{54}$Fe was utilized to measure the electron-induced knockout of protons with high initial momentum. 
In this experiment, the additional eight $f\frac{7}{2}$-neutrons in ${}^{48}$Ca contribute to a 10\% increase in high-momentum protons over $^{40}$Ca, while the additional six mostly-$f\frac{7}{2}$ protons in $^{54}$Fe contribute to a 50\% increase in high-momentum protons over $^{48}$Ca.
However, according to the analysis of Ref.~\cite{Duer:2018} a 40\% increase would be expected in high-momentum protons in ${}^{48}$Ca compared to ${}^{40}$Ca because the mass of the isotopes used (${}^{12}$C, ${}^{27}$Al, ${}^{56}$Fe, and ${}^{208}$Pb) also increases with the increase of neutron excess, and the experiment cannot separately extract the effect of neutron excess on the increase of high-momentum protons.

None of the existing models based on the Similarity Renormalization Group (SRG)~\cite{tropiano:2024high}, the Low-Order Correlation Operator Approximation (LCA)~\cite{ryckebusch:2019isospin}, and various phenomenological quantum-pairing models~\cite{vanhalst:2012quantifying, Lane:1964nuclear}, can simultaneously describe the increases in high-momentum protons observed in the CaFe experiment. 
While these models yield comparable results for the high-momentum ratio of ${}^{48}$Ca to ${}^{40}$Ca reported in Ref.~\cite{Nguyen2026}, none of the models coincide with the implied trend reported in Ref.~\cite{Duer:2018} for a high-momentum ratio of ${}^{54}$Fe to ${}^{40}$Ca. The CaFe experiment implies that intra-orbital pairing is much stronger than inter-orbital pairing. Moreover, the Class collaboration emphasized the connection between the long-range nuclear shell structure and the short-range correlation which emerges as a high-momentum component in the momentum distributions.

The dispersive optical model (DOM) generates the nucleon self-energy within the framework of Green's function theory~\cite{Exposed!}, constrained by experimental observables (both below and above the Fermi energy), and provides a complete picture of LRC and SRC correlations.
It also addresses the fraction of high-momentum protons and neutrons in the momentum distribution~\cite{Mahzoon:2014,Dickhoff:2017,atkinson2020dispersive}. These many-body correlations evolve with energy from LRC near the Fermi energy to substantially localized SRC away from the Fermi energy. A new extension of the DOM~\cite{Ramon:2026c} has proven to be very successful in explaining the extracted experimental values of the ${}^{40}$Ca spectral function $S(E_{m},p_{m})$ by incorporating a shrinkable geometry of the volume imaginary part.

In this letter, we present the high-momentum ratios of ${}^{48}$Ca to ${}^{40}$Ca and ${}^{54}$Fe to ${}^{48}$Ca calculated using the DOM with the recent extension~\cite{Ramon:2026c} which shows excellent quantitative agreement with the experimental values obtained from the CaFe experiment analysis~\cite{Nguyen2026}. Unlike other nuclear models, the nonlocal version of the DOM successfully forges a link between the domains of nuclear reactions and structure~\cite{Mahzoon:2014}. Moreover, the recent extension~\cite{Ramon:2026c} of DOM provides a better description of the self-energy and spectral functions, leading to a quantitatively correct description of the momentum distribution.

In the many-body Green's function formalism, the irreducible self-energy, $\Sigma^*(\bm{r},\bm{r}';E)$, is a complex one-body potential which, in principle, is comprised of an infinite set of Feynman diagrams describing the propagation of an interacting nucleon through a nucleus based on a Hamiltonian containing relevant two- and three-body interactions~\cite{Exposed!}. In the DOM formalism, this complex one-body potential is parametrized as an optical potential which naturally extends to negative (bound) energies within the Green's function framework as first employed by Mahaux and Sartor~\cite{Mahaux91}. 
 The analytic structure of the nucleon self-energy allows one to utilize a dispersion relation, which relates the real part of the self-energy at a given energy to a dispersion integral of its imaginary part over all energies. The energy-independent correlated Hartree-Fock (HF) contribution~\cite{Exposed!} is removed by employing a subtracted dispersion relation with the Fermi energy used as the subtraction point~\cite{Mahaux91}
  \begin{align}
    \mathrm{Re}\ \Sigma^*(\bm{r},\bm{r}';E) &= \mathrm{Re}\
    \Sigma^*(\bm{r},\bm{r}';\varepsilon_F) \label{eq:dispersion} \\ -
    \mathcal{P}\int_{\varepsilon_F}^{\infty} \!\! \frac{dE'}{\pi}&\mathrm{Im}\
    \Sigma^*(\bm{r},\bm{r}';E')\left[\frac{1}{E-E'}-\frac{1}{\varepsilon_F-E'}\right] \nonumber
    \\ + \mathcal{P} \! \int_{-\infty}^{\varepsilon_F} \!\!
    \frac{dE'}{\pi}&\mathrm{Im}\
    \Sigma^*(\bm{r},\bm{r}';E')\left[\frac{1}{E-E'}-\frac{1}{\varepsilon_F-E'}\right],
    \nonumber      
 \end{align}
 where $\bm{r}$ also implies relevant discrete quantum numbers and $\varepsilon_F$ is the average Fermi energy which separates the particle and hole domains, 
 \begin{align*}
 \varepsilon_F = \frac{1}{2}\left(E_0^{A+1}-E_0^{A-1}\right),
 \end{align*}
 and $E_0^{A\pm1}$ represent the ground state energies of the $A\pm1$ nucleus~\cite{Exposed!}. 
 
Earlier work to include ground-state properties in the DOM demonstrated that it was not possible to reproduce the charge density without including a nonlocal imaginary part~\cite{Dickhoff:2010}.
We therefore continue to implement a nonlocal representation of the self-energy following
 Ref.~\cite{Mahzoon:2014} where $\Sigma_{\text{HF}}(\bm{r},\bm{r'}) = \mathrm{Re}\
    \Sigma^*(\bm{r},\bm{r}';\varepsilon_F) $ and the imaginary part 
 $\mathrm{Im}\ \Sigma(\bm{r},\bm{r'};E)$ are parametrized, and Eq.~\eqref{eq:dispersion} generates the energy dependence of the
 real part. The HF term consists of a volume term, spin-orbit term,  and a wine-bottle shape consistent with a microscopic analysis~\cite{Brida11}. The imaginary self-energy consists of volume, surface, and spin-orbit terms. 
 Nonlocality is represented using the Gaussian form as proposed in Ref.~\cite{Perey:1962}. 
 More details can be found in Ref.~\cite{atkinson2020dispersive}. The parameterization is presented in the supplement~\cite{Ramon:2026ExtensionSupp}. 
 
To use the DOM self-energy for predictions, the parameters are fit through a weighted $\chi^2$ minimization of available elastic differential cross section data ($\frac{d\sigma}{d\Omega}$), analyzing power data ($A_\theta$),  reaction cross sections ($\sigma_r$), total cross sections ($\sigma_t$), charge density ($\rho_{\text{ch}}$), energy levels ($\varepsilon_{\ell j}$), particle number, the root-mean-square charge radius ($R_\mathrm{ch}$), and the energy of the ground state. The scattering calculations are performed in a partial wave basis (specifying orbital and total angular momentum with $\ell$ and $j$, respectively) using $\Sigma_{\ell j}^*(r,r';E)$ as an optical potential in the framework of $R$-matrix theory~\cite{Baye:2010}.
 All calculations are done in a Lagrange basis with 30 mesh points, where Legendre polynomials with a matching radius of 12 fm are used for scattering calculations and Laguerre polynomials are used for bound-state calculations~\cite{Baye_review,Baye:2010}.
  
We employ the Dyson equation to obtain the Green's function, $G_{\ell j}(r,r';E)$, from the DOM self-energy~\cite{Exposed!}.
 The particle number, binding energy, and charge density are all obtained from the hole spectral function which corresponds to the imaginary part of the Green's function, 
 \begin{equation}
    S^{(P,N)}_{\ell j}(r,r';E) = \frac{1}{\pi}\mathrm{Im}\ G^{(P,N)}_{\ell j}(r,r';E) ,
    \label{eq:spec}
 \end{equation}
where $P$ and $N$ denote proton and neutron, respectively.

The spectral function in momentum space for a given $\ell j$ can be obtained by employing the double Fourier-Bessel transformation of the imaginary part of the propagator expressed in coordinate space,
 \begin{eqnarray}
    \label{eq:momentum_space_S}
    S^{(P,N)}_{lj}(k;E)
    &=& \frac{2}{\pi^2}\int_{0}^{\infty} dr r^{2} \int_{0}^{\infty} dr' r'^{2} \\
    &\times&j_{\ell}(kr)~j_{\ell}(kr')~\mathrm{Im}~G_{\ell j}^{(P,N)}(r,r';E), \nonumber
 \end{eqnarray}
The momentum distributions can be obtained by integrating spectral functions over missing energy, 
 \begin{equation}
    \label{eq:momentum_distribution}
    n^{(P,N)}(k) =\sum_{\ell j} (2j+1) \int_{-\infty}^{\varepsilon_F}dE\ S^{(P,N)}_{\ell j}(k;E).
 \end{equation}
 
While the focus of the present paper is to study high-momentum components in the DOM framework, we note that the DOM is capable of describing the removal of valence protons near the Fermi energy.
Indeed, the DOM self-energy generates all the ingredients for an accurate distorted-wave impulse approximation of the  $(e,e'p)$ cross section,
as demonstrated in Refs.~\cite{Atkinson:2018,Atkinson:2019,Atkinson:2024,Ramon:2026B}.
The accuracy of these results quantitatively dictate the need for the admixture of single-particle strength below the Fermi energy just as the elastic scattering data indicate where some of the removed strength now resides~\cite{Dussan:2014}.

\begin{figure}[t]
    \includegraphics[width=\columnwidth]{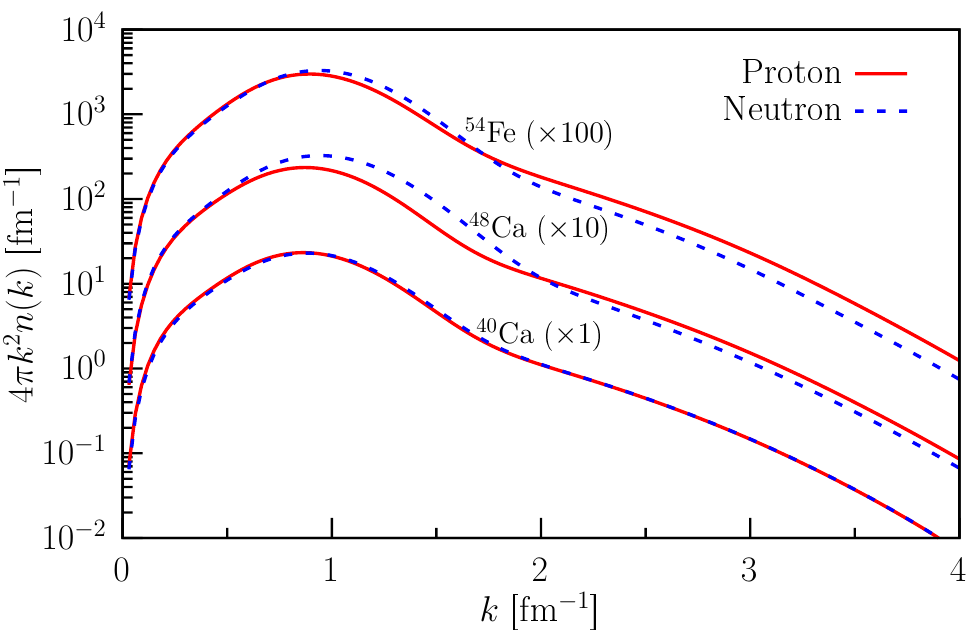}
    \caption{DOM-calculated proton and neutron momentum distributions with the implementation of energy dependence to the geometry of the volume imaginary part of the self-energy below the Fermi energy~\cite{Ramon:2026c}. Red solid lines and blue dashed lines indicate the proton and neutron momentum distributions, respectively.}
   \label{fig:kdist_combined}
\end{figure}

For the first time, we report the DOM-calculated momentum distribution of ${}^{54}$Fe along with the simultaneous description of nuclear reactions and structure (see the supplement) of ${}^{54}$Fe. The proton and neutron momentum distributions, $n(k)$ in Fig.~\ref{fig:kdist_combined} were calculated using Eq.~\eqref{eq:momentum_distribution} in the DOM framework for ${}^{40}$Ca, ${}^{48}$Ca, and ${}^{54}$Fe. Here, the region of $n(k)$ above $k >1.52~\mathrm{fm^{-1}}~(300~\mathrm{MeV/c})$ is considered the high-momentum region~\cite{Hen:2014momentum, Duer:2018, Duer:2019direct}.  
For ${}^{40}$Ca, the proton and neutron momentum distributions (see Fig.~\ref{fig:kdist_combined}) are essentially identical to each other throughout the whole momentum range, because ${}^{40}$Ca is a doubly magic symmetric $(N=Z)$ nucleus with the highest mostly occupied orbital $1d\frac{3}{2}$ for both protons and neutrons. In the doubly magic asymmetric $(N\neq Z)$ nucleus, ${}^{48}$Ca, 8 additional neutrons mostly occupy the $f\frac{7}{2}$ orbital and change the proton and neutron momentum distribution. In Fig.~\ref{fig:kdist_combined}, the neutron momentum distribution of ${}^{48}$Ca shows an enhancement in the momentum range $k=0.5$ to $2.0 \mathrm~{fm^{-1}}$ due to the additional 8 neutrons mostly in the $f\frac{7}{2}$ orbit beyond the ${}^{40}$Ca core. Despite the addition of 8 neutrons in this orbital in ${}^{48}$Ca, the proton $n(k)$ exceeds the neutron $n(k)$ for $k\gtrsim2.0~\mathrm{fm^{-1}}$ while incorporating a thinner neutron skin $(0.122~\mathrm{fm})$~\cite{Ramon:2026}. This behavior in $n(k)$ is expected according to the np dominance picture established by high-momentum knockout measurements~\cite{Duer:2018, Egiyan:2006Measurement} and \textit{ab initio} calculations of asymmetric matter~\cite{Rios:2009,Rios:2014}.
The addition of 6 protons mostly to the $f\frac{7}{2}$ orbital in going from ${}^{48}$Ca to ${}^{54}$Fe increases proton high-momentum fraction for $k\gtrsim1.75~\mathrm{fm^{-1}}$, and this enhancement in proton $n(k)$ becomes significancy larger compared to the high-momentum protons in ${}^{48}$Ca demonstrating that the $f\frac{7}{2}$ protons and $f\frac{7}{2}$ neutrons are highly correlated (paired).
\begin{figure}[b]
    \includegraphics[width=\columnwidth]{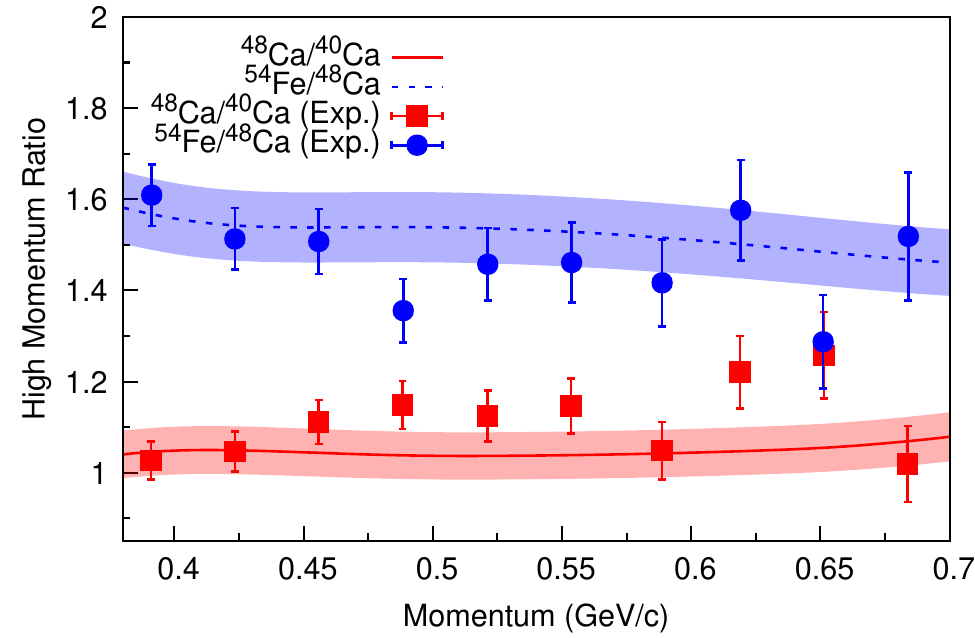}
    \caption{High momentum ratios. Ratio of high-momentum components of protons for ${}^{48}$Ca/${}^{40}$Ca are represented with solid red squares and ${}^{54}$Fe/${}^{48}$Ca are indicated by solid blue circles. The solid red line with faded red band represents the DOM-calculated high-momentum ratio for ${}^{48}$Ca/${}^{40}$Ca, and the blue dashed line with faded blue band shows the DOM-calculated high-momentum ratio of ${}^{54}$Fe/${}^{48}$Ca. Experimental data are from Ref.~\cite{Nguyen2026}.}
   \label{fig:high_momentum_ratio}
\end{figure}
Fig.~\ref{fig:high_momentum_ratio} shows the high momentum ratios of ${}^{48}$Ca/${}^{40}$Ca with solid squares (red)  and  ${}^{54}$Fe/${}^{48}$Ca with solid circles (blue) obtained from the CaFe experiment~\cite{Nguyen2026}. The experimental data covers the range of momenta from $k=0.39$ to $0.68~\mathrm{GeV/c}$ which corresponds to $k=0.1.98$ to $3.45~\mathrm{fm^{-1}}$. The DOM-calculated proton high momentum ratios are shown in full (red) and dashed (blue) lines with shaded bands of errors for ${}^{48}$Ca/${}^{40}$Ca and  ${}^{54}$Fe/${}^{48}$Ca, respectively, using the momentum distribution shown in Fig.~\ref{fig:kdist_combined}. 
The DOM error bands are ultimately related to the errors in the experimental data used in constraining the self-energy (see Refs.~\cite{Mahzoon:2017,atkinson2020dispersive} for the uncertainty quantification).
Figure~\ref{fig:high_momentum_ratio} shows that the DOM calculations reproduce the experimentally measured high-momentum ratios in the CaFe nuclei with remarkable agreement for the entire experimental momentum range.

This analysis of the CaFe nuclei employs the improved version of the DOM, which includes an energy-dependent geometry of the volume imaginary part of the self-energy below the Fermi energy. This new implementation has proven to be very successful, providing a consistent description of experimental values of the spectral function at high missing momenta and energies~\cite{Ramon:2026c}. All the necessary details of this new extension are given in Ref.~\cite{Ramon:2026c} along with the physically correct description of the proton-neutron high-momentum distribution of ${}^{48}$Ca while achieving the small neutron skin of 0.122~fm.
\begin{figure}[b]
    \includegraphics[width=\columnwidth]{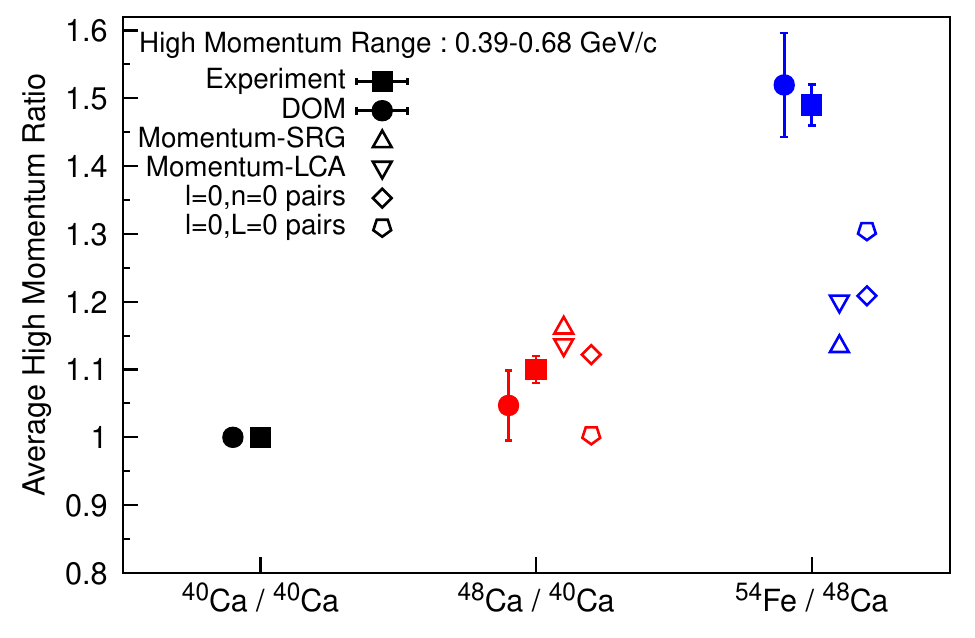}
    \caption{Average of the high-momentum ratios. Solid squares indicate the values from CaFe experiment. Solid circles are DOM calculated ratios. Hollow symbols indicate different theoretical predictions as indicated in the legend. black, red and blue corresponds to ratios ${}^{48}$Ca/${}^{40}$Ca, ${}^{48}$Ca/${}^{40}$Ca and ${}^{54}$Fe/${}^{48}$Ca. The symbols associated to theoretical values are shifted slightly left and right of the experimental values for clarity. The average of the DOM ratios were calculated for the experimental range $0.39-0.68$ GeV of data available in Ref.~\cite{Nguyen2026}.}
   \label{fig:high_momentum_ratio_average}
\end{figure}
Since almost all high-momentum contributions originate from the SRC pairs, 
the shape of the momentum distribution, $n(k)$, is independent of the characteristics of the nucleus~\cite{Duer:2018,Duer:2019direct,Carlson:2015,Wiringa:2014} (also confirmed in Fig.\ref{fig:kdist_combined}). Consequently, the ratios of high momentum fraction of ${}^{48}$Ca to ${}^{40}$Ca and ${}^{54}$Fe to ${}^{48}$Ca are approximately constant across the measured range of missing momenta ($k=0.1.98$ to $3.45~\mathrm{fm^{-1}}$). As a result, the weighted averages of the experimental data points are also a good indicator to compare with theoretical results.

In Fig.~\ref{fig:high_momentum_ratio_average}, the solid squares are the experimental averages of the high-momentum ratios compared with DOM results represented by solid circles. The DOM calculated average high-momentum ratios are $1.05~\pm~0.05$ and $1.52~\pm~0.08$ for ${}^{48}\mathrm{Ca}$/${}^{40}\mathrm{Ca}$ and ${}^{54}\mathrm{Fe}$/${}^{48}\mathrm{Ca}$ and show very good agreement with the corresponding experimental values of $1.10~\pm~0.02$ and $1.49~\pm~0.03$~\cite{Nguyen2026}, while all other theoretical results indicated by different hollow symbols  do not explain both the ratios simultaneously.

Theoretical models based on the similarity renormalization
group (SRG)~\cite{tropiano:2024high} calculate single-nucleon momentum distributions by combining phenomenological Woods-Saxon single-particle wave functions with operators evolved via the SRG. Although SRG-based results for ${}^{48}$Ca/${}^{40}$Ca are minimally higher than the corresponding experimental value, the ${}^{54}$Fe/${}^{48}$Ca experimental ratio is substantially underestimated.
Low-order correlation operator approximation (LCA) calculations~\cite{ryckebusch:2019isospin} also give good agreement with the experimental ${}^{48}$Ca/${}^{40}$Ca high momentum ratio while considerably underestimating ${}^{54}$Fe/${}^{48}$Ca ratio. 
The phenomenological quantum-pairing model in Ref.~\cite{vanhalst:2012quantifying} yielded values comparable to LCA-based calculated values (see Fig.\ref{fig:high_momentum_ratio_average}) for both ratios, where they quantify the SRC by computing the number of IPM nucleon pairs and triples, which reveal the behavior beyond the mean-field treatment. In particular, numbers of pn, pp, and nn SRC pairs with zero relative orbital momentum state $(n=0,l=0)$ were calculated. Another quantum-pairing model was introduced in Ref.~\cite{Nguyen2026} avoiding inter-orbital pairing by the requirement of SRC pairs to have zero relative and total angular momentum $(l=0, L=0)$. This model underestimated the experimental value of the ratio, ${}^{48}$Ca/${}^{40}$Ca, and shows an improved result for ${}^{54}$Fe/${}^{48}$Ca compared to results based on SRG, LCA, and the quantum-pairing model with $(n=0,l=0)$. However, it only predicted an increase of $30\%$ high momentum proton content in ${}^{54}$Fe relative to ${}^{48}$Ca, where the experimentally high-momentum in ${}^{54}$Fe was increased by $50\%$.

Compared to other theoretical models, the extended DOM is the only framework that simultaneously describes both experimental ratios, ${}^{48}$Ca/${}^{40}$Ca and ${}^{54}$Fe/${}^{48}$Ca. The extended DOM thus provides an excellent description of SRC physics, the observed high-momentum components, and their evolution with nuclear shell structure. We emphasize that no fit was made to the experimental ratios, so the constraints provided by other experimental data (see supplement) were sufficient to obtain agreement.
The success of the DOM to describe high-momentum physics suggests that future experiments should attempt to determine the complete spectral functions including their energy dependence. This could help to clarify the transition to nuclear-matter type conclusions about high-momentum results using the DOM for heavier nuclei.

This work was supported by the U.S. National Science Foundation under grants PHY-2207756 and PHY-2512895.

\bibliographystyle{apsrev4-1}
\bibliography{CaFe}

\end{document}